\documentclass{ws-ijmpd}

\usepackage[super,compress]{cite}

\usepackage{color}
\usepackage{ulem}

\begin{document}

\markboth{Satoshi Tsuchida and Masaki Mori}
{The electron plus positron spectrum from annihilation of Kaluza--Klein dark matter in the Galaxy}

%
\catchline{}{}{}{}{}
%

\title{The electron plus positron spectrum from annihilation of Kaluza--Klein dark matter in the Galaxy}

\author{Satoshi Tsuchida$^{*}$}

\address{Department of Physics, Osaka City University, 3--3--138 Sugimoto, Sumiyoshi--ku, Osaka City, Osaka, 558--8585, Japan \\
stsuchida88@gmail.com}

\author{Masaki Mori}

\address{Department of Physical Sciences, Ritsumeikan University, Kusatsu 525--8577, Shiga, Japan}

\maketitle

\begin{history}
\received{Day Month Year}
\revised{Day Month Year}
\end{history}

\begin{abstract}%
The lightest Kaluza--Klein particle (LKP), which appears in
the theory of universal extra dimensions,
is one of good candidates for cold dark matter (CDM).
When LKP pairs annihilate around the center of the Galaxy where CDM is concentrated,
there are some modes which produce electrons and
positrons as final products, and we categorize them into two components.
One of them is the ``Line'' component, which directly annihilates into electron--positron pair.
Another one is the ``Continuum'' component, which consists of secondarily produced electrons and positrons via some decay modes.
Before reaching Earth, directions of electrons and positrons are randomized by the Galactic magnetic field,
and their energies are reduced by energy loss mechanisms.
We assume the LKP is in the mass range from 300~GeV to 1500~GeV.
We calculate the electron plus positron spectrum after propagation in the Galactic halo to Earth,
and we analyze the resulting spectrum and positron fraction.
We also point out that the energy dependence of observed positron fraction is well reproduced
by the mixture of ``line'' and ``continuum'' components.
We can fit the electron plus positron spectrum and the positron fraction by assuming
appropriate boost factors describing dark matter concentration in the Galactic halo.
However, it is difficult to explain both the electron plus positron spectrum and the positron fraction by a single boost factor,
if we take account of observational data obtained by AMS--02 only.
\end{abstract}


\ccode{PACS numbers: 95.35.+d; 98.70.Sa}

\section{Introduction}

The fact that most of the matter in the Universe consists of non--baryonic dark matter~\cite{Freese1999}
is supported further by the Planck observational data~\cite{Ade2014},
and dark matter should be made of particles which do not exist in the standard model of particle physics.
Recent observations of cosmic positron excess~\cite{Adriani2010, Ackermann2012, Aguilar2013} could
be explained by secondarily produced positrons in annihilation of dark matter particles in the Galactic halo,
as is discussed by many authors (see, e.g. Refs.~\cite{Kopp2013, Feng2014, Calore2015}).
Among various candidates of dark matter, the lightest Kaluza--Klein particle (LKP),
predicted in the theory of universal extra dimensions (UED)~\cite{Appelquist2001, Cheng2002a, Servant2003},
is unique since there would be a characteristic edge structure 
in the cosmic electron plus positron spectrum near the LKP mass as Cheng $et~al$.~\cite{Cheng2002b} predicted.
The edge structure was calculated by Moiseev $et~al$.~\cite{Moiseev2007} for Fermi--LAT detection,
but, at least in the energy range below 1000 GeV,
such structure has not established so far (see, e.g. Refs.~\cite{Ackermann2012, Ting2013, Aguilar2014}).
On the other hand, above 1000 GeV,
the observational data are still limited, so the characteristic
structure could be observed in near--future missions.
For example, the Calorimetric Electron Telescope (CALET), which is a Japanese--led detector and is a fine resolution calorimeter
for cosmic--ray observation installed on the International Space Station in August 2015, started
exploring the energy range up to 20 TeV for electrons and positrons~\cite{Torii2015}.

In this paper, we calculate the electron and positron spectrum 
and the positron fraction from LKP annihilation in the Galactic halo 
including the effects of propagation, and compare the results with recent measurements. 
We found the energy dependence of the positron fraction is well explained
by the mixture of ``line'', which consists of electron--positron pairs directly produced by annihilation,
and ``continuum'' component, which consists of secondarily produced electrons and positrons.
Also the total electron plus positron spectrum can be explained by taking account of the LKP contribution
assuming appropriate values for the boost factor describing concentration of dark matter in the Galactic halo.
Then we discuss the constraints on the boost factor.

\section{Production of electrons and positrons}

In UED assuming only one extra dimension,
the extra dimension is compactified with radius $ R $,
and the LKP mass, which we denote as $ m_{B^{(1)}} $, is inversely proportional to $ R $.
The relevant mass for the LKP ranges from a few 100~GeV to 1000~GeV~\cite{Bergstrom2005a},
if we assume the LKP contributes significantly to cold dark matter.
More recently, progress in estimating the relic density indicate it could be as heavy as 1.5 TeV~\cite{Belanger2011}.
Experimentally, recent LHC results indicate the LKP mass lighter than 950 GeV is disfavored~\cite{Atlas2015}. 
Here, we vary the LKP mass from 300 GeV to 1500 GeV, and we analyze the
electron plus positron spectrum and the positron fraction to be observed at Earth from LKP annihilation.

\begin{figure}[t]
  \begin{center}
    \includegraphics[width=\columnwidth]{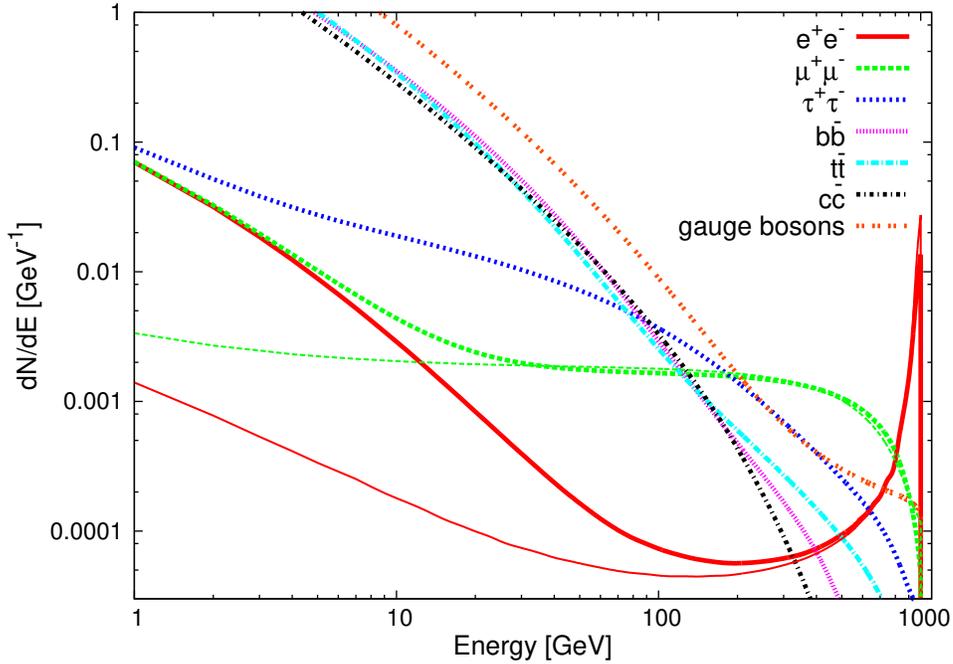}
    \caption{(Color Online).
      The ``continuum'' and ``line'' positron spectra from LKP annihilations for $ m_{B^{(1)}} = 1000 $~GeV \cite{Cirelli2011, Ciafaloni2011}.
      The patterned lines correspond to the positron spectra
      per annihilation via muon pairs, tauon pairs, quark pairs, and gauge bosons, respectively.
      The solid line corresponds to the line component.
      The thick lines include the electroweak corrections, and the thin lines do not include those corrections.}
    \label{fig:continuum}
  \end{center}
\end{figure}

When LKP pairs annihilate, there are some modes which produce electrons and positrons as final products,
and we categorize them into two components.
One of them is a ``line'' component, which consists of electron--positron pairs directly produced by annihilation,
and gives rise to edge structure near the LKP mass after propagating in the Galactic halo to Earth.
Another is a ``continuum'' component, which consists of secondarily produced electrons and positrons
via muon pairs, tauon pairs, quark pairs, and gauge bosons produced by LKP annihilation.
The branching ratios are given as follows:
20{\%} for charged leptons,
11{\%} for up--type quarks,
0.7{\%} for down--type quarks,
1{\%} for charged gauge bosons,
and 0.5{\%} for neutral gauge bosons, respectively~\cite{Servant2003, Hooper2003}.
We use the spectra for line and continuum components given by Cirelli $et~al$. \cite{Cirelli2011}
and Ciafaloni $et~al$. \cite{Ciafaloni2011},
which are shown in Fig. \ref{fig:continuum} assuming 100{\%} branching ratios for each component,
where the solid line indicates the line spectrum and 
patterned lines show the continuum spectra
from muon pairs, tauon pairs, quark pairs ($b$, $t$, $c$),
and gauge bosons, respectively.
Note that the line spectra shows a tail toward lower energies
due to final state interactions.
For comparison, the positron spectra without electroweak corrections are shown in thin lines
for the line spectrum and the continuum spectrum for muon pairs.
One can see the electroweak correction affects the spectra in the lower energy region~\cite{Cirelli2011, Ciafaloni2011}.

The spectra for line and continuum components shown in Fig.~\ref{fig:continuum} are those just after pair annihilation,
and we have to take account of the effects of propagation in the Galactic halo to Earth,
such as diffusion and energy loss processes.
For this purpose, we follow the Green function approach given by Moskalenko and Strong~\cite{Moskalenko1999},
assuming the ``Isothermal model'' as the halo profile for reference.
In addition, we should include a ``boost factor'', $ B_{f} $,
which describes the signal enhancement from dark matter annihilation in the Galactic halo~\cite{Prada2004}.
$ N $--body simulation study given by Navarro, Frenk and White (NFW)~\cite{Navarro1996}, for example, indicates a large $ B_{f} $.
The boost factor $ B_{f} $ is defined as following expression
\begin{eqnarray}
  \label{eq:booost}
  B_{f} &=& B_{\rho} \times B_{\sigma v}  \nonumber \\
    &=& \left( \frac{ \langle {\rho}^{2} (l) \rangle_{\Delta V} }{ \langle {\rho}_{0}^{2} (l) \rangle_{\Delta V} } \right)
    \left( \frac{ \langle \sigma v \rangle }{ 3 \times 10^{-26} ~{\rm{cm^{3} s^{-1}}} } \right)_{\Delta V}
\end{eqnarray}
where $ {\rho}_{0} $ is a local dark matter density estimated as 0.43~$ { \rm{ GeV / cm^{3} } } $
for ``Isothermal'' halo model \cite{Moskalenko1999},
and $ 3 \times 10^{-26}~{ \rm{cm^{3} s^{-1}}} $
is the typical cross section for cold dark matter annihilation \cite{Bergstrom2009}.

The values of $B_{f}$ could be determined to fit the observed positron fraction
as discussed extensively to interpret the ``anomaly'' reported by PAMELA~\cite{Adriani2010}.
For example, Cirelli $et~al$. discussed the values of $B_{f}$ for each annihilation mode \cite{Cirelli2009}.
However,
the energy dependence of the positron fraction
observed by AMS--02 with more statistics~\cite{Aguilar2013}
is not well explained if only the line (or continuum) component is taken into account.
We will show later that we can fit it well by considering both the line and the continuum components
assuming an appropriate boost factor $B_{f}$.

\section{The effect of propagation}

Charged particles, such as electrons and positrons, produced by LKP annihilation around the center of the Galaxy
change their direction randomly by the irregular component in the Galactic magnetic field,
and lose their energies by bremsstrahlung in interstellar matter before reaching Earth.
Thus, the observational electron plus positron fluxes have different shapes from initial ones.
The effects of propagation are studied by Moskalenko and Strong \cite{Moskalenko1999},
and we calculate the modulated flux by using their results given as Green functions.

The positron flux is given by \cite{Moskalenko1999};
\begin{eqnarray}
  \label{eq:posifluxfeng}
  \frac{ d {\Phi}_{ e^{+} } }{ d {\Omega} dE } = \langle \sigma v \rangle B_{f}
  \left( \frac{ {\rho}_{0} }{ m_{B^{(1)}} } \right)^{2} \sum_{i} B_{i} \int d{\epsilon} \frac{ dN_{i} }{ d{\epsilon} } g \left( \epsilon , E \right)~{\rm{cm^{-2} s^{-1} sr^{-1} GeV^{-1}}}
\end{eqnarray}
where $ B_{i} $ is a branching ratio for each particle,
and $ {\rho}_{0} $ is the local dark matter density.
The annihilation cross section $ \langle \sigma v \rangle $
to yield the significant relic density of cold dark matter
is the order of $ 3 \times 10^{-26}~{{\rm{cm^{3}s^{-1}}}}$ \cite{Bergstrom2009}.

The Green function, $ g ( \epsilon , E ) $, is defined as
\begin{eqnarray}
  \label{eq:greenfunc}
  g \left( \epsilon , E \right) = \frac{ 10^{25} }{ E^{2} } 10^{ a ( \log_{10} E )^{2} + b ( \log_{10} E ) + c } \theta \left( \epsilon - E \right)
\end{eqnarray}
where $E$ is the observed energy in GeV and the parameters $ a $, $ b $ and $ c $ are tabulated in Ref.~\cite{Moskalenko1999},
which gives the spectra of electron and positron after propagation, for monochromatic energy ($\epsilon$) injection.

In addition, we should take account of the effects of solar modulation
in the low energy region below 10~GeV.
The magnetic field of the Sun is the source of the observed modulation
of the Galactic cosmic rays.
Solar modulation is dominant on low energy particles,
and affects on spectral shape for cosmic rays.
In the force field approximation, the differential flux of particles of mass $m$ and charge $Ze$, ${\Phi}(E)$,
reaching Earth with energy $E$ is related to the interstellar flux, ${\Phi}(E_{IS})$, as
\begin{eqnarray}
  \label{eq:solarmod}
  {\Phi} (E) = \frac{ E^{2} - m^{2} }{ E_{IS}^{2} - m^{2} } {\Phi} (E_{IS})
\end{eqnarray}
where $E_{IS}$ is the energy in interstellar space and related to $E$ as
$ E = E_{IS} - \vert Z \vert \phi $, and $ \phi $ is a solar modulation potential
\cite{Stanev2003}.

The line component is approximately in the form of $ \delta $--function
before propagation in the Galactic halo.
However, its spectrum after propagation in the Galactic halo to Earth extends to lower energies
caused by the effects of diffusion and energy loss processes.
\begin{figure}[t]
  \begin{center}
    \includegraphics[width=\columnwidth]{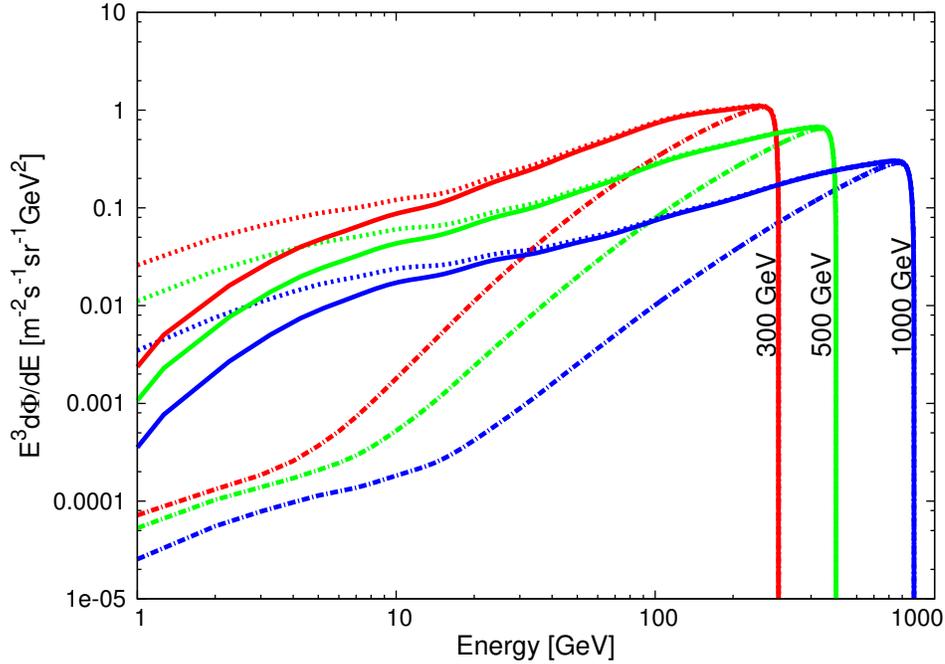}
    \caption{(Color Online).
      The spectra of electrons plus positrons from LKP annihilation after propagation for three assumed LKP masses. 
      The dot--dashed lines show the spectrum without solar modulation for the ``line'' component only,
      the dotted lines show the total spectrum from LKP annihilation (``continuum'' plus ``line'') without solar modulation,
      and the solid lines show the total spectrum including effect of solar modulation
      with the solar modulation potential $ {\phi} = 735~{\rm{MV}} $.
      We assume the boost factor $ B_{f} = 1 $.}
    \label{fig:line35}
  \end{center}
\end{figure}
In addition to the line component, we also calculate the continuum component.
The continuum component has a broad spectrum extending to lower energies when it is produced by LKP annihilation as shown in Fig.~\ref{fig:continuum}.
Then we calculated the spectrum after propagation using the Green function similarly to the case of the line component.
These results for the case of the isothermal halo density profile are shown in Fig.~\ref{fig:line35},
where the dot--dashed lines show the spectrum for the line component only without solar modulation,
the dotted lines show the total spectrum from LKP annihilation (continuum plus line) without solar modulation,
and solid lines show the total spectrum including effect of solar modulation
assuming a solar modulation potential $ {\phi} = 735~{\rm{MV}} $ \cite{Yuan2013} for three assumed LKP masses.
This figure indicates the continuum component becomes dominant in lower energies,
and it is larger by two orders of magnitude around 10~GeV than the line component.

\begin{figure}[h]
  \begin{center}
    \includegraphics[width=\columnwidth]{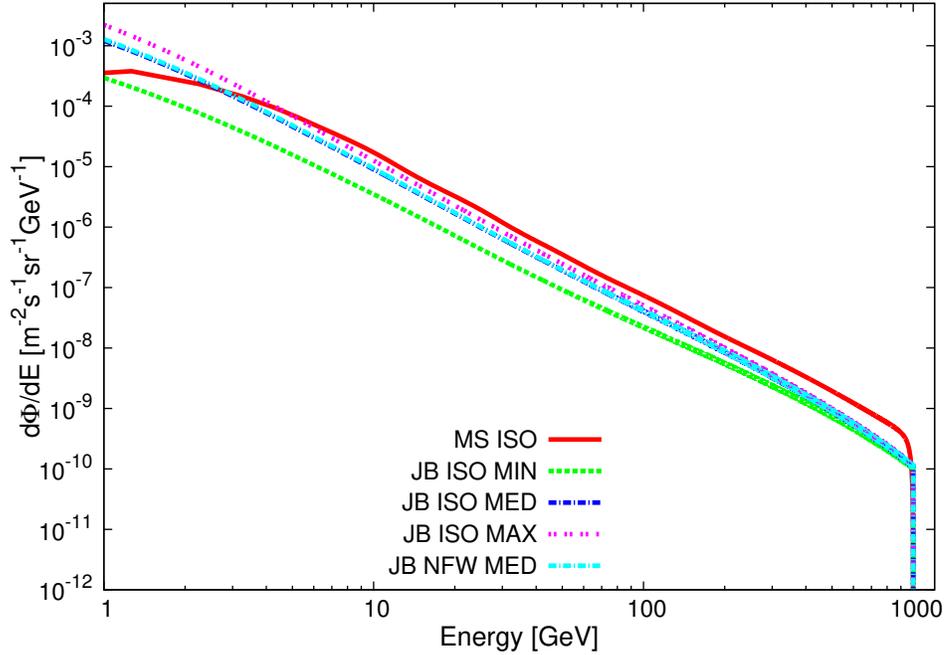}
    \caption{(Color Online).
      The spectra of electrons plus positrons from LKP annihilation after propagation for $ m_{B^{(1)}} = 1000 $~GeV.
      The solid line corresponds to solid line in Fig.~\ref{fig:line35} based on calculation given by
      Moskalenko and Strong, ``MS''.
      Other lines show the spectra based on calculation given by J. Buch, ``JB'',
      for each halo density and propagation model (see text).
      We assume the boost factor $ B_{f} = 1 $.}
    \label{fig:e3fluxfigure}
  \end{center}
\end{figure}

We checked our calculation by comparison with a similar calculation given by Buch $et~al$.~\cite{Buch2015}.
In Fig.~\ref{fig:e3fluxfigure} the spectra of electrons plus positrons from LKP annihilation after propagation
are plotted for $ m_{B^{(1)}} = 1000 $~GeV.
Plots indicated by ``JB'' are given in Ref.~\cite{Buch2015} for various combination of parameters:
``NFW'' or ``ISO'' (isothermal) for halo density models, 
``MIN'', ``MED'', and ``MAX'' for halo propagation models
(here ``MF1'' for the Galactic magnetic field model is assumed). 
Detailed descriptions of these parameter sets are given in Refs.~\cite{Buch2015, Donato2004}.
These plots indicate that two halo density models give almost the same spectra,
while different halo propagation models affect the spectra especially
in the lower energy region.
Also shown in Fig.~\ref{fig:e3fluxfigure} indicated by ``MS'' is our calculation 
using the Green function approach based on calucation by
Moskalenko and Strong~\cite{Moskalenko1999}
(corresponding to the solid line in Fig.~\ref{fig:line35}).
The MS spectrum is larger than JB by a factor of about three just below the LKP mass,
however the overall shape in the higher energy region ($\gtrsim 10$~GeV) of the plots looks similar.
We will use only the data in this energy region observed by AMS--02 as discussed below,
and those models (MS and JB) give similar results.
We also note that the parameters for the Green function used in MS are given for the dark matter mass up to 1000~GeV,
but we would like to discuss the behavior of the boost factor when the dark matter mass is heavier than 1000~GeV.
Thus, in the following discussion,
we will calculate the electron plus positron spectrum and the positron fraction to
set constraints on the boost factor assuming the propagation models of JB only
to avoid complicate discussion and readers' confusion.

\section{Discussion}

So far, we have calculated the electron plus positron spectrum from LKP annihilation, which we denote as $ F_{\rm{LKP}} (E)$.
Now, we compare the result of calculation with recent measurements to discuss possible constraints on the boost factor.

Yuan and Bi calculated the cosmic--ray secondary (hereafter ``conventional'') electron plus positron spectrum~\cite{Yuan2013}, which we denote as $F_{\rm{Conv}} (E)$.
We note that a normalization factor, $ c_{e^{+}} $, is included in their calculation to fit the observational data,
so we treat that the overall normalization factor of $ F_{\rm{Conv}} (E)$ as a free parameter,
which reduces a degree of freedom by unity.

Then, we assume the total electron plus positron spectrum as follows
\begin{eqnarray}
  \label{eq:model_flux}
  {\Phi}^{e^{\pm}} (E) = B_{f} \times F_{\rm{LKP}} (E) + C \times F_{\rm{Conv}} (E)
\end{eqnarray}
where a coefficient $C$ is a normalization factor, which corresponds to $c_{e^{+}}$ in Ref.~\cite{Yuan2013}.
Here we employ the least--squares method to obtain appropriate values for $B_{f}$ and $C$ to fit the AMS--02 data~\cite{Aguilar2014}.
The goodness of the fit can be tested by the sum
\begin{eqnarray}
  \label{eq:chisqvalue}
  {\chi}^{2} = \sum_{i} \frac{ \left( {\rm{data}} - {\rm{model}} \right)^{2} }{ {\sigma}_{\rm{data}}^{2} }
\end{eqnarray}
where ``data'' means the data points, ``model'' is given by Eq.~(\ref{eq:model_flux}), and
we assume $ {\sigma}_{\rm{data}} = \sigma_{\rm{stat}}+\sigma_{\rm{syst}}$ as
a conservative estimate of the error of the data points as the worst case, instead of 
$ {\sigma}_{\rm{data}} =\sqrt{\sigma_{\rm{stat}}^2+\sigma_{\rm{syst}}^2}$ as is usually assumed.
Here, $ {\sigma}_{\rm{stat}} $ and $ {\sigma}_{\rm{syst}} $ are the statistic and systematic errors for the electron plus positron spectrum
quoted by the AMS--02 collaboration~\cite{Aguilar2014}.
The index $i$ runs the data points in the energy range between about 30 and 1000~GeV.
A number of degrees of freedom is 29 (=~a number of data points (31) minus unknown parameter (2)).
Thus, ${\chi}^{2} < 49.6$ is required to be consistent with the AMS--02 data at 99{\%} confidence level.

\begin{figure}[t]
  \begin{center}
    \includegraphics[width = \columnwidth]{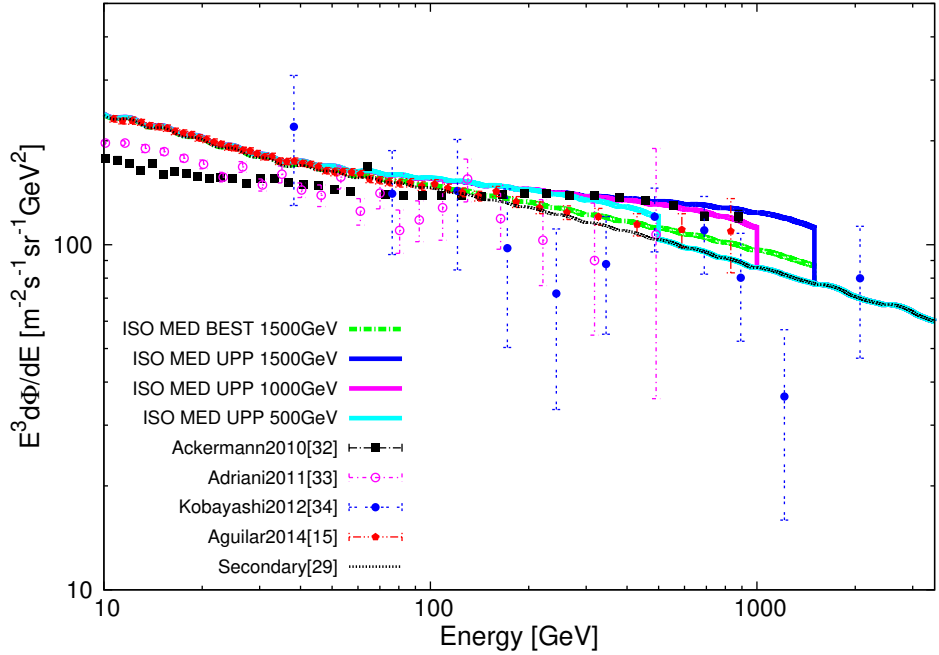}
    \caption{(Color Online).
      The electron plus positron spectra assuming several LKP masses, $m_{B^{(1)}}$,
      and boost factors, $B_{f}$.
      The dot--dashed line shows the spectrum fit to AMS--02 data with the best fit boost factor, $B_{f} = 130$,
      for isothermal halo model and MED propagation model for $m_{B^{(1)}} = 1500$~GeV.
      Thick solid lines are the spectral fits to the AMS--02 data with upper--limit boost factor,
      $ B_{f} = 80,~240,~465$ for $m_{B^{(1)}} = 500,~1000,~1500$~GeV, respectively.
      Also shown are the recent observational data~\cite{Kobayashi2012, Ackermann2010, Adriani2011, Aguilar2014},
      and an adjusted (factor $C=1.2$ is multiplied) prediction spectrum by the cosmic--ray secondary calculation (thin dotted line) \cite{Yuan2013}.}
    \label{fig:e3fluxexpfigure}
  \end{center}  
\end{figure}

We calculate ${\chi}^{2}$ values for various parameter sets, where we vary the factor $C$ from 1 to 2 in 0.1 step
and the boost factor $B_{f}$ from 0 to 1500 in 5 step.
The results of this calculation show that we have no parameter set to fit AMS--02 data at 99{\%} confidence level unless the factor $C=1.2$,
even if we take any value for the boost factor (from 0 to 1500).
When we set $C$ equals to 1.2, we can obtain a range of the values of $B_{f}$ under the condition ${\chi}^{2} < 49.6$ for each LKP mass,
halo density, and propagation model, which are given in Table~\ref{table:flux_ams}.
The result for the case of $B_{f}=0$ implies that we can explain the total electron plus positron spectrum
observed by AMS--02 without LKP contribution.
\begin{table}[t]
  \tbl{The values of the boost factor assuming Isothermal (NFW) halo model and for each propagation model with the best fit (B), lower (L) and upper (U) limit to AMS--02 obeservational spectrum.} 
  { \begin{tabular}{ c | c c c | c c c | c c c } \toprule 
      LKP mass      &        \multicolumn{3}{|c|}{MIN}        &         \multicolumn{3}{|c|}{MED}       &        \multicolumn{3}{|c}{MAX}         \\
      $ [{\rm{GeV}}] $  & $ B_{f} $ L & $ B_{f} $ B & $ B_{f} $ U & $ B_{f} $ L & $ B_{f} $ B & $ B_{f} $ U & $ B_{f} $ L & $ B_{f} $ B & $ B_{f} $ U \\ \hline
      500       &    0 (0)    &   35 (35)   &  125 (125)  &    0 (0)    &   20 (20)   &   80 (80)   &    0 (0)    &   20 (20)   &    70 (65)  \\
      1000       &    0 (0)    &  105 (105)  &  390 (390)  &    0 (0)    &   70 (65)   &  240 (240)  &    0 (0)    &   60 (55)   &   205 (200) \\
      1200       &    0 (0)    &  145 (145)  &  530 (535)  &    0 (0)    &   90 (90)   &  325 (320)  &    0 (0)    &   80 (75)   &   275 (265) \\
      1500       &    0 (0)    &  210 (210)  &  775 (775)  &    0 (0)    &  130 (130)  &  465 (460)  &    0 (0)    &  110 (105)  &   390 (375) \\  \hline \hline
  \end{tabular} }
  \label{table:flux_ams}
\end{table}
The resulting total electron plus positron spectra by using some parameter sets given in Table~\ref{table:flux_ams} are shown in Fig.~\ref{fig:e3fluxexpfigure}.
The dot--dashed line shows the spectrum with the best fit boost factor, $B_{f} = 130$, for isothermal halo model and MED propagation model for $m_{B^{(1)}} = 1500$~GeV.
Thick solid lines are the spectral fits to the AMS--02 data with upper--limit boost factors,
where $ B_{f} = 80,~240,~465$ for $m_{B^{(1)}} = 500,~1000,~1500$~GeV, respectively.
Thin dashed line shows a conventional spectrum~\cite{Yuan2013} multiplied by $C=1.2$.
This figure shows the edge structure will be clearer for heavier LKP mass,
and we can expect to differentiate such structures in the higher energy region by future observations
with higher sensitivity.

\begin{figure}[htbp]
     \begin{center}
      \includegraphics[width=\columnwidth]{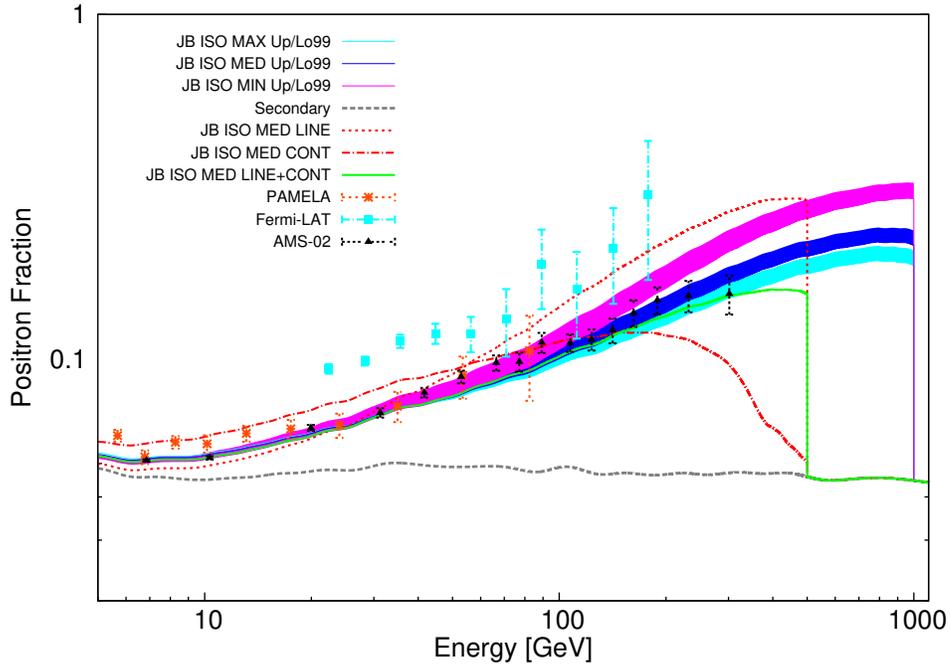}
      \caption{(Color Online).
	The positron fraction expected from LKP annihilation for each propagation model
	compared with recent measurements \cite{Adriani2010, Ackermann2012, Aguilar2013},
	and a prediction by the cosmic--ray secondary 
	electron and positron calculation (thin dashed line) \cite{Yuan2013}.
	Thick solid lines show the range of expected positron fraction by using obtained upper and lower limit on $B_{f}$ at
	99{\%} confidence level for each propagation model assuming $m_{B^{(1)}}= 1000$~GeV.
	Thin solid line shows the expected positron fraction including not only the line component but also the continuum component, 
        thin dotted line shows that without the continuum component, and
	thin dot--dashed line shows that without the line component for MED propagation model and $m_{B^{(1)}} = 500$~GeV.
      }
      \label{fig:posifrafigure}
    \end{center}
\end{figure}

Now, we discuss the constraints on the boost factor $B_{f}$
based on the positron fraction observed by AMS--02~\cite{Aguilar2013}.
The positron fraction, $ e^{+} / \left( e^{-} + e^{+} \right) $, for the conventional spectrum,
$ f_{\rm{Conv}} (E) $, depends on energy but is generally smaller than 0.1~\cite{Yuan2013}.
On the other hand, the LKP pair annihilation creates the same number of electrons and positrons,
so the positron fraction for the LKP spectrum, $ f_{\rm{LKP}} $, always equals to 0.5.
Then, the total positron fraction is given by
\begin{eqnarray}
  \label{eq:posifra}
  {\rm{Positron \ Fraction}} = \frac{ F_{\rm{LKP}} (E) \times B_{f} \times f_{\rm{LKP}} + F_{\rm{Conv}} (E) \times f_{\rm{Conv}} (E) }{ F_{\rm{LKP}} (E) \times B_{f}+ F_{\rm{Conv}} (E) },
\end{eqnarray}
Then, we employ the least--squares method to obtain the appropriate value for $B_{f}$
to fit the AMS--02 data~\cite{Aguilar2013}.
In a similar way with a analysis of the total electron plus positron spectrum,
the goodness of the fit can be tested by the sum as Eq.~(\ref{eq:chisqvalue}).
In this case, ``model'' is given by Eq.~(\ref{eq:posifra}), and
the data points ``data'',  $ {\sigma}_{\rm{stat}} $ and $ {\sigma}_{\rm{syst}} $ are the statistic and systematic errors for positron fraction
quoted by the AMS--02 collaboration~\cite{Aguilar2013}.
The index $i$ runs the data points in the energy range between 10 and about 400~GeV.
A number of degrees of freedom is 42 ($=43-1$).
Thus, ${\chi}^{2} < 66.2$ is required to be consistent with the AMS--02 data at 99{\%} confidence level.
We obtain a range of the values of $B_{f}$ under the condition ${\chi}^{2} < 66.2$ for each LKP mass, halo density and propagation models.
With this prescription, we calculate the positron fractions for several assumed LKP masses and propagation models
as a function of energy to fit the AMS--02 data \cite{Aguilar2013} as shown in Fig.~\ref{fig:posifrafigure}.
One can see the observed data can be fit well by adding the LKP flux with an appropriate boost factor,
and the energy dependence of the data can be fit better if we include not only the ``line'' but also the ``continuum'' component.
Also for that case using the MED propagation model seems to give a better fit.
The values of the boost factor, $B_{f}$, derived as above for each LKP mass are given in Table~\ref{table:posifra_ams}.
\begin{table}[t]
  \tbl{The value of boost factor assuming Isothermal (NFW) halo model and each propagation model with the best fit (B), lower (L) and upper (U) limit to AMS02 obeservational data (Positron Fraction).}
    { \begin{tabular}{ c | c c c | c c c | c c c } \toprule
       LKP mass      &        \multicolumn{3}{|c|}{MIN}        &         \multicolumn{3}{|c|}{MED}       &        \multicolumn{3}{|c}{MAX}         \\
   $ [{\rm{GeV}}] $  & $ B_{f} $ L & $ B_{f} $ B & $ B_{f} $ U & $ B_{f} $ L & $ B_{f} $ B & $ B_{f} $ U & $ B_{f} $ L & $ B_{f} $ B & $ B_{f} $ U \\ \hline
           500       &  343 (343)  &  393 (393)  &  444 (444)  &  152 (149)  &  178 (173)  &  203 (199)  &  116 (101)  &  135 (118)  &  154 (135)  \\
          1000       &  983 (984)  & 1150 (1150) & 1320 (1320) &  467 (456)  &  519 (506)  &  571 (555)  &  353 (308)  &  390 (341)  &  428 (375)  \\
          1200       & 1286 (1287) & 1500 (1500) & 1717 (1716) &  627 (612)  &  679 (661)  &  732 (710)  &  473 (415)  &  510 (446)  &  547 (477)  \\
          1500       & 1775 (1775) & 2058 (2058) & 2199 (2199) &  909 (894)  &  937 (913)  &  964 (931)  &    No Fit   &    No Fit   &    No Fit   \\ \hline \hline
    \end{tabular} }
  \label{table:posifra_ams}
\end{table}
If we assume $ {\rm{LKP~mass}} = 1500 $~GeV and MAX propagation model,
there would be no allowed value for $ B_{f} $ to be consistent with the positron fraction observed by AMS--02
at 99{\%} confidence level.

From the results given in Tables~\ref{table:flux_ams} and~\ref{table:posifra_ams},
we observe the {\it{lower}} limit on $ B_{f} $ to be consistent with the positron fraction
may be larger than the {\it{upper}} limit on $ B_{f} $ to be consistent with the electron plus positron spectrum.
Although we can fit to the positron fraction and the electron plus positron spectrum separately by adding LKP contribution,
the required boost factors differ significantly and it is difficult to explain both the electron plus positron spectrum and the positron fraction at the same time,
if we only take account of observational data obtained by AMS--02.
However, there remains rather large differences in the total electron plus positron spectrum among various experiments,
so it is a bit too early to conclude whether LKP survives as a relevant dark matter candidate or not.

It is important to compare our results with those given in previous studies.
The detectability of gamma rays from dark matter annihilation has been discussed in many literatures.
For instance, Bergstr{\"o}m $et~al$.~\cite{Bergstrom2012} discussed the relation between the mass of dark matter and cross section,
and predicted an upper limit of the cross section about $9 \times 10^{-28}~{\rm{cm^{3}~s^{-1}}}$
assuming 1000~GeV dark matter mass for HESS--II observation.
Assuming $130 \times 10^{-6}$~pb for the cross section for annihilation into photon pairs,
this upper limit corresponds to about $B_{f}=230$.
In our calculation, by assuming $m_{B^{(1)}}=1000$~GeV, Isothermal halo model and MED halo propagation based on the JB propagation,
we obtain $B_{f} =$~0--240 (to explain the total spectrum) or 467--571 (to explain the positron fraction),
which fit to AMS--02 observational data at 99{\%} confidence level.
Thus, the upper limit imposed by gamma--ray observation is comparable with our result.
In addition, $B_{f}$ is also calculated in the case for proton--antiproton observation.
Cholis $et~al$.~\cite{Cholis2012} discussed the value of the boost factor based on the data of proton--antiproton flux obtained by the PAMELA observation.
They pointed out the value of $B_{f}$ should be ${\cal{O}}(100)$ for 1000~GeV dark matter mass,
which is not contradiction with our result.

In the energy region around or higher than 1000~GeV, the measurements suffer rather large statistical uncertainties.
Thus, we hope on--going (CALET~\cite{Torii2011} and DAMPE~\cite{Wu2015})
and near--future missions with better sensitivity will clarify
the existence (or non--existence) of the LKP dark matter.

\section{Conclusion}

We investigated the cosmic electron and positron spectra from LKP annihilation
taking account of propagation effects in the Galaxy.
We paid particular attention to the calculation of the ``continuum'' emission,
which is a secondary product of LKP annihilation,
in addition to the ``line'' component directly produced by annihilation.
The result shown in Fig.~\ref{fig:line35} indicates the ``continuum'' component dominates
over the ``line'' component in the low energy region after propagation in the Galactic halo to Earth,
and changes the shape of the positron fraction, as shown in Fig. \ref{fig:posifrafigure}.
We also consider the spectra for different halo density and propagation models.
The results mean that the difference of halo density models do not affect on the spectra effectively,
but the choice of halo propagation models change the shape of spectra in the lower energy region, as shown in Fig.~\ref{fig:e3fluxfigure}.

We estimated the value of the boost factor
to enhance the halo density in the Galactic center region
by using the electron plus positron spectrum and the positron fraction measurement by AMS--02,
which is given in Tables~\ref{table:flux_ams} and~\ref{table:posifra_ams}.
The results of our calculation imply that
while the addition of the LKP component gives good fits to
the total electron plus positron spectrum and the positron fraction
with appropriate boost factors estimated for each case,
but these two boost factors are not consistent each other,
if we only take account of the AMS--02 observation.
However, considering the rather large experimental uncertainties
we should not conclude that whether LKP can be a good candidate of cold dark matter now.

If the characteristic structure in electron plus positron flux is observed in near future,
we may conclude dark matter is made of LKP.
It would be a conclusive evidence for the existence of extra dimensions.

\section*{Acknowledgments}

We would like to thank Mr. Akihiko Kawamura and Dr. Fumihiro Matsui
for useful discussions and helpful comments.
We also would like to thank Marco Cirelli and his collaborators
for supplying numerical data of their calculation.

\end{document}